# Comparing Erlang distribution and Schwinger mechanism on transverse momentum spectra in high energy collisions

Li-Na Gao and Fu-Hu Liu[1]

*Institute of Theoretical Physics, Shanxi University, Taiyuan, Shanxi 030006, China*

**Abstract:** We study the transverse momentum spectra of $J/\psi$ and $\Upsilon$ mesons by using two methods: the two-component Erlang distribution and the two-component Schwinger mechanism. The results obtained by the two methods are compared and found to be in agreement with the experimental data of proton-proton (*pp*), proton-lead (*p*-Pb), and lead-lead (Pb-Pb) collisions measured by the LHCb and ALICE Collaborations at the large hadron collider (LHC). The related parameters such as the mean transverse momentum contributed by each parton in the first (second) component in the two-component Erlang distribution and the string tension between two partons in the first (second) component in the two-component Schwinger mechanism are extracted.

## 1. Introduction

In the last century, scientists predicted that a new state of matter could be produced in relativistic heavy-ion (nucleus-nucleus) collisions or could exist in quark stars owing to high temperature and high density [1-3]. The new matter is named the quark-gluon plasma (QGP) or quark matter. This prediction makes the research of high energy collisions developing rapidly. A lot of physics researchers devote in research the mechanisms of particle productions and the properties of QGP formation. Because of the reaction time of the impacting system being very short, people could not make a direct measurement for the collision process. So, only by researching the final state particles, people can presumed the evolutionary process of collision system. For this reason, people proposed many models to simulate the process of high energy collisions [4].

The transverse momentum (mass) spectra of particles in final state are an important observation. They play one of major roles in high energy collisions. Other quantities which also play major roles include, but not limited to, pseudorapidity (or rapidity) distribution, azimuthal distribution (anisotropic flow), particle ratio, various correlations, etc. [5]. Presently, many formulas such as the standard (Fermi-Dirac, Bose-Einstein, or Boltzmann) distribution [6], the Tsallis statistics [7-11], the Tsallis form of standard distribution [11], the Erlang distribution [12], the Schwinger mechanism [13-17], and others are used in describing the transverse momentum spectra. It is expected that the excitation degree (effective temperature and kinetic freeze-out temperature), radial flow velocity, and other information can be obtained by analyzing the particle transverse momentum spectra.

---
[1] E-mail: fuhuliu@163.com; fuhuliu@sxu.edu.cn

In this paper, we use two methods, the two-component Erlang distribution and the two-component Schwinger mechanism, to describe the transverse momentum spectra of $J/\psi$ and $\Upsilon$ mesons produced in proton-proton (*pp*), proton-lead (*p*-Pb), and lead-lead (Pb-Pb) collisions measured by the LHCb and ALICE Collaborations at the large hadron collider (LHC) [18-21]. The related parameters such as the mean transverse momentum contributed by each parton in the first (second) component in the two-component Erlang distribution and the string tension between two partons in the first (second) component in the two-component Schwinger mechanism are extracted.

## 2. Formulism

We assume that the basic impacting process in high energy collisions is binary parton-parton collision. We have two considerations on the description of violent degree of the collision. A consideration is the mean transverse momentum contributed by each parton. The other one is the string tension between two partons. The former consideration can be studied in the framework of Erlang distribution. The latter one results in the Schwinger mechanism. Considering the wide transverse momentum spectra in experiments, we use the two-component Erlang distribution and the two-component Schwinger mechanism. Generally, the first component describes the region of low transverse momentum, and the second one describes the high transverse momentum region.

*Firstly*, we consider the two-component Erlang distribution. Let $\langle p_t \rangle_1$ and $\langle p_t \rangle_2$ denote the mean transverse momentums contributed by each parton in the first and second component respectively. Each parton is assumed to contribute an exponential transverse momentum ($p_t$) spectrum. For the *j*th (*j*=1 and 2) parton in the *i*th component, we have the distribution

$$f_{ij}(p_{tj}) = \frac{1}{\langle p_t \rangle_i} \exp\left(-p_{tj}/\langle p_t \rangle_i\right). \tag{1}$$

The transverse momentum ($p_T$) in final state is $p_T = p_{t1} + p_{t2}$. The transverse momentum distribution in final state is the folding of $f_{i1}(p_{t1})$ and $f_{i2}(p_{t2})$. Considering the contribution ratio $k_{E1}$ of the first component, we have the (simplest) two-component Erlang distribution to be

$$f(p_T) = k_{E1} \int_0^{p_T} f_{11}(p_{t1}) f_{12}(p_T - p_{t1}) dp_{t1} + (1 - k_{E1}) \int_0^{p_T} f_{21}(p_{t1}) f_{22}(p_T - p_{t1}) dp_{t1}$$

$$= \frac{k_{E1} p_T}{\langle p_t \rangle_1^2} \exp\left(-p_T/\langle p_t \rangle_1\right) + \frac{(1 - k_{E1}) p_T}{\langle p_t \rangle_2^2} \exp\left(-p_T/\langle p_t \rangle_2\right). \tag{2}$$

In the Monte Carlo method, let $R_{1,2}$ denote random numbers in [0,1]. For the *i*th component, we have

$$p_T = p_{t1} + p_{t2} = -\langle p_t \rangle_i (\ln R_1 + \ln R_2). \tag{3}$$

We would like to point out that the folding of multiple exponential distributions with the same parameter results in the ordinary Erlang distribution which is not used in the present work. The Monte Carlo method performs a simpler calculation for the folding.

*Secondly*, we consider the two-component Schwinger mechanism. Let $\kappa_1$ and $\kappa_2$ denote the string tensions between the two partons in the first and second component respectively, $K_1 \equiv \kappa_1/\pi$, $K_2 \equiv \kappa_2/\pi$, and $m_0$ denotes the rest mass of a parton. According to [13-17], for the *j*th (*j*=1 and 2) parton in the given string in the *i*th component, we have the distribution

$$f_{ij}(p_{tj}) = C_0(K_i) \exp\left[-(p_{tj}^2 + m_0^2)/K_i\right]$$

$$= C_0(K_i) \exp(-m_0^2/K_i) \exp(-p_{tj}^2/K_i) = \frac{1}{\sqrt{K_i \pi}} \exp(-p_{tj}^2/K_i), \tag{4}$$

where

$$C_0(K_i) = 1 \bigg/ \int_0^\infty \exp\left[-(p_{tj}^2 + m_0^2)/K_i\right] dp_{tj} = \frac{1}{\sqrt{K_i \pi}} \exp(m_0^2/K_i)$$

is the normalization constant. Considering the contribution ratio $k_{S1}$ of the first component, we have the two-component Schwinger mechanism

$$f(p_T) = k_{S1} \int_0^{p_T} f_{11}(p_{t1}) f_{12}(p_T - p_{t1}) dp_{t1} + (1 - k_{S1}) \int_0^{p_T} f_{21}(p_{t1}) f_{22}(p_T - p_{t1}) dp_{t1}$$

$$= k_{S1} C_0^2(K_1) \int_0^{p_T} \exp\left\{-\left[p_{t1}^2 + (p_T - p_{t1})^2 + 2m_0^2\right]/K_1\right\} dp_{t1}$$

$$+ (1 - k_{S1}) C_0^2(K_2) \int_0^{p_T} \exp\left\{-\left[p_{t1}^2 + (p_T - p_{t1})^2 + 2m_0^2\right]/K_2\right\} dp_{t1}$$

$$= \frac{k_{S1}}{K_1 \pi} \int_0^{p_T} \exp\left\{-\left[p_{t1}^2 + (p_T - p_{t1})^2\right]/K_1\right\} dp_{t1}$$

$$+ \frac{1 - k_{S1}}{K_2 \pi} \int_0^{p_T} \exp\left\{-\left[p_{t1}^2 + (p_T - p_{t1})^2\right]/K_2\right\} dp_{t1}. \tag{5}$$

In the Monte Carlo method, let $R_{3-6}$ denote random numbers in [0,1]. For the *i*th component, we have

$$p_T = p_{t1} + p_{t2} = \sqrt{K_i} \left[\sqrt{-\ln R_3} \cos(2\pi R_4) + \sqrt{-\ln R_5} \cos(2\pi R_6)\right]. \tag{6}$$

## 3. Results

The transverse momentum spectra, $d^2\sigma(J/\psi)/dp_T dy$, of $J/\psi$ mesons produced in different data samples in *pp* collision at center-of-mass energy $\sqrt{s} = 7$ TeV are shown in Figure 1, where $\sigma$ and $y$ denote the cross section and rapidity respectively. The symbols represent the experimental data measured by the LHCb Collaboration [18] in different rapidity ranges and scaled by different amounts marked in the panels. The dashed and solid curves are our results calculated by using the two-component Erlang distribution and the two-component Schwinger mechanism respectively. From Figures 1(a) to 1(d), the data samples are prompt $J/\psi$ with no polarisation, $J/\psi$ from $b$ with no polarisation, prompt $J/\psi$ with full transverse polarisation, and prompt $J/\psi$ with full longitudinal polarisation, respectively. The values of free parameters and $\chi^2$ per degree of freedom ($\chi^2/dof$) are listed in Table 1. One can see that the two methods describe the experimental data of the LHCb Collaboration.

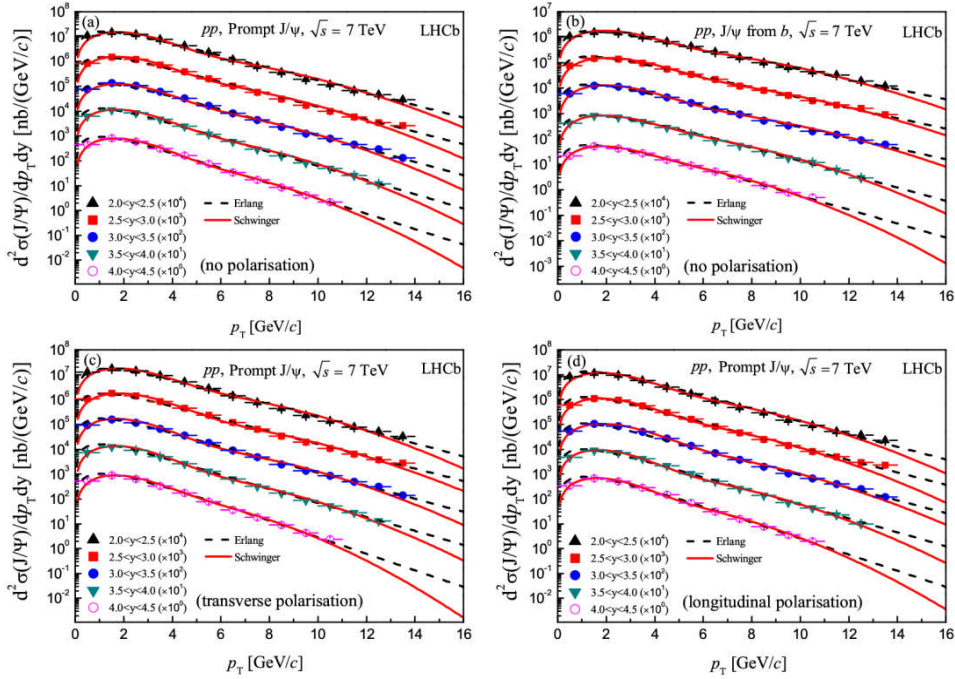

Figure 1: Transverse momentum spectra of (a) prompt $J/\psi$ with no polarisation, (b) $J/\psi$ from $b$ with no polarisation, (c) prompt $J/\psi$ with full transverse polarisation, and (d) prompt $J/\psi$ with full longitudinal polarisation in $pp$ collision at $\sqrt{s}=7$ TeV. The symbols represent the experimental data measured by the LHCb Collaboration [18] in different rapidity ranges and scaled by different amounts marked in the panels. The dashed and solid curves are our results calculated by using the two-component Erlang distribution and the two-component Schwinger mechanism respectively.

Figures 2(a) and 2(b) present the transverse momentum spectra, $d^2\sigma(J/\psi)/dp_T dy$, of $J/\psi$ mesons produced in data samples, prompt $J/\psi$ and $J/\psi$ from $b$, in $pp$ collision at $\sqrt{s}=8$ TeV respectively. Figures 3(a), 3(b), and 3(c) present the transverse momentum spectra, $B^{nS}d^2\sigma^{nS}/dp_T dy$, of $\Upsilon(1S)$, $\Upsilon(2S)$, and $\Upsilon(3S)$ mesons produced in the same collision respectively, where $n=1, 2,$ and 3 correspond to the three mesons respectively, and $B$ denotes the branching ratio. The symbols represent the experimental data of the LHCb Collaboration [19] and the curves are our results. Figure 4 is similar to Figure 2, but it presents the results in $p$-Pb collisions at center-of-mass energy per nucleon pair $\sqrt{s_{NN}} = 5$ TeV, and the experiment data of the LHCb collaboration are taken from ref. [20]. The values of free parameters and $\chi^2/dof$ corresponding to Figures 2-4 are listed in Table 1. Similar conclusion obtained from Figure 1 can be obtained from Figures 2-4, though some of fitting results are approximately in agreement with the data.

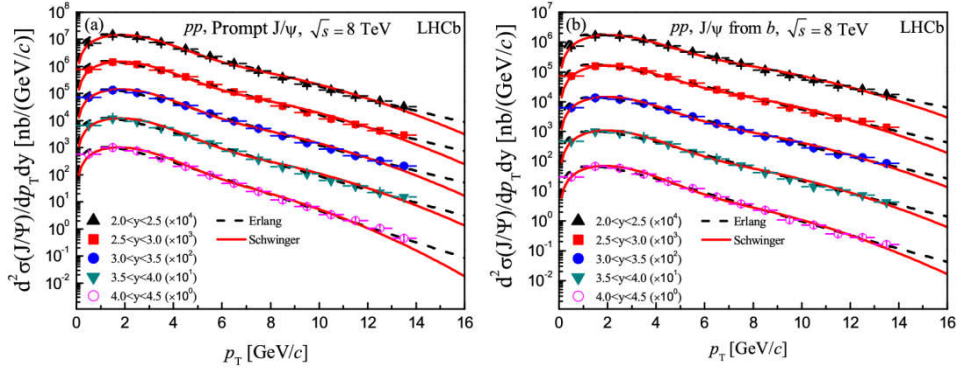

Figure 2: Transverse momentum spectra of (a) prompt $J/\psi$ and (b) $J/\psi$ from $b$ in $pp$ collision at $\sqrt{s} = 8$ TeV. The symbols represent the experimental data of the LHCb Collaboration [19]. The dashed and solid curves are our results calculated by using the two-component Erlang distribution and the two-component Schwinger mechanism respectively.

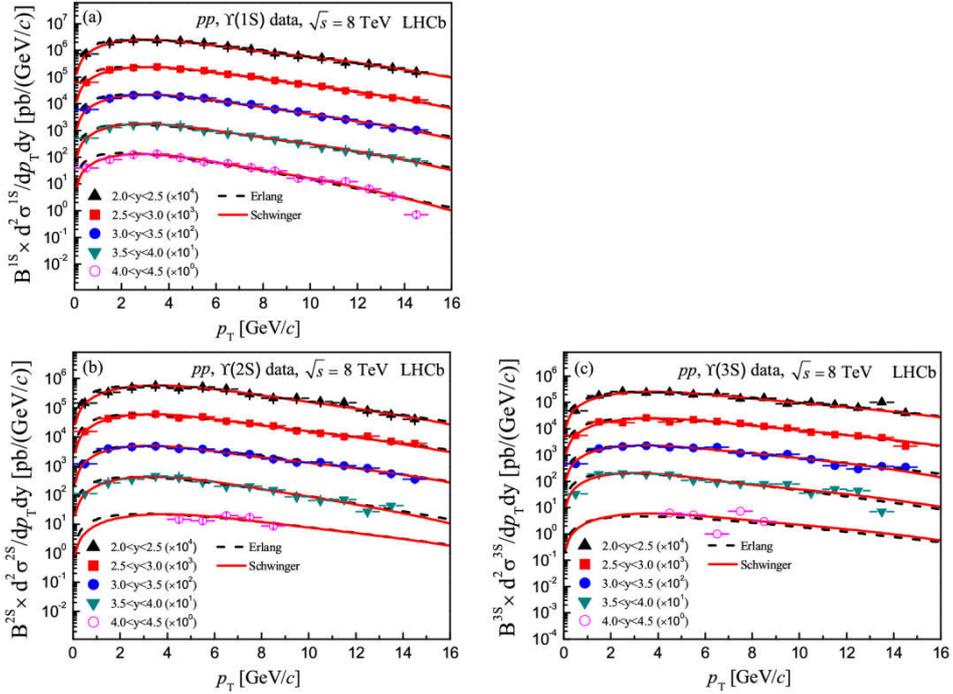

Figure 3: Transverse momentum spectra of (a) $\Upsilon(1S)$, (b) $\Upsilon(2S)$, and (c) $\Upsilon(3S)$ mesons produced in $pp$ collision at $\sqrt{s} = 8$ TeV. The symbols represent the experimental data of the LHCb Collaboration [19]. The dashed and solid curves are our results calculated by using the two-component Erlang distribution and the two-component Schwinger mechanism respectively.

Table 1: Values of parameters and $\chi^2/dof$ corresponding to the curves in Figures 1-5.

| Figure | Type | Two-component Erlang | | | | Two-component Schwinger | | | |
|---|---|---|---|---|---|---|---|---|---|
| | | $k_{E1}$ | $\langle p_t \rangle_1$ (GeV/c) | $\langle p_t \rangle_2$ (GeV/c) | $\chi^2/dof$ | $k_{S1}$ | $\kappa_1$ (GeV/fm) | $\kappa_2$ (GeV/fm) | $\chi^2/dof$ |
| Figure 1(a) | $2.0 < y < 2.5$ | 0.540±0.062 | 1.008±0.132 | 1.553±0.086 | 0.519 | 0.813±0.027 | 11.70±0.88 | 54.30±3.67 | 1.814 |
| | $2.5 < y < 3.0$ | 0.545±0.083 | 1.008±0.107 | 1.533±0.082 | 0.846 | 0.810±0.023 | 11.50±1.20 | 50.00±3.32 | 2.318 |
| | $3.0 < y < 3.5$ | 0.552±0.102 | 0.978±0.124 | 1.473±0.078 | 0.775 | 0.817±0.029 | 11.42±1.24 | 47.82±3.98 | 1.882 |
| | $3.5 < y < 4.0$ | 0.557±0.096 | 0.887±0.093 | 1.373±0.070 | 0.927 | 0.830±0.020 | 10.85±1.10 | 44.00±3.82 | 0.544 |
| | $4.0 < y < 4.5$ | 0.564±0.093 | 0.878±0.120 | 1.283±0.100 | 0.913 | 0.830±0.020 | 10.43±1.12 | 38.00±4.23 | 0.256 |
| Figure 1(b) | $2.0 < y < 2.5$ | 0.563±0.084 | 1.183±0.087 | 1.957±0.094 | 0.226 | 0.713±0.023 | 13.80±3.12 | 65.00±7.40 | 0.880 |
| | $2.5 < y < 3.0$ | 0.582±0.063 | 1.120±0.093 | 1.886±0.112 | 0.940 | 0.778±0.032 | 15.90±3.78 | 69.50±7.90 | 0.492 |
| | $3.0 < y < 3.5$ | 0.605±0.078 | 1.118±0.100 | 1.838±0.098 | 0.922 | 0.786±0.037 | 14.60±3.65 | 66.00±7.50 | 0.474 |
| | $3.5 < y < 4.0$ | 0.650±0.068 | 1.300±0.092 | 1.570±0.087 | 0.571 | 0.700±0.028 | 12.80±3.07 | 49.80±6.80 | 0.291 |
| | $4.0 < y < 4.5$ | 0.647±0.100 | 1.180±0.140 | 1.498±0.153 | 0.105 | 0.657±0.023 | 11.50±3.05 | 40.30±6.30 | 0.218 |
| Figure 1(c) | $2.0 < y < 2.5$ | 0.610±0.077 | 1.180±0.093 | 1.538±0.102 | 0.738 | 0.785±0.025 | 11.90±2.75 | 50.00±5.90 | 2.545 |
| | $2.5 < y < 3.0$ | 0.640±0.080 | 1.008±0.100 | 1.580±0.087 | 0.705 | 0.860±0.036 | 12.80±2.93 | 55.00±6.30 | 1.255 |
| | $3.0 < y < 3.5$ | 0.650±0.065 | 0.953±0.127 | 1.500±0.076 | 0.711 | 0.853±0.032 | 11.00±2.67 | 49.30±6.10 | 1.353 |
| | $3.5 < y < 4.0$ | 0.656±0.072 | 0.902±0.118 | 1.404±0.092 | 1.049 | 0.855±0.027 | 10.80±2.70 | 45.00±5.70 | 0.523 |
| | $4.0 < y < 4.5$ | 0.725±0.082 | 1.010±0.120 | 1.255±0.085 | 0.707 | 0.776±0.023 | 10.00±2.63 | 33.20±5.20 | 0.708 |
| Figure 1(d) | $2.0 < y < 2.5$ | 0.715±0.062 | 1.120±0.140 | 1.603±0.117 | 0.842 | 0.793±0.038 | 12.00±3.80 | 50.00±5.60 | 2.816 |
| | $2.5 < y < 3.0$ | 0.640±0.074 | 1.008±0.093 | 1.580±0.082 | 1.171 | 0.796±0.043 | 12.00±4.20 | 48.00±5.80 | 4.194 |
| | $3.0 < y < 3.5$ | 0.596±0.054 | 0.947±0.073 | 1.530±0.078 | 1.108 | 0.847±0.032 | 13.00±3.60 | 52.00±5.20 | 0.789 |
| | $3.5 < y < 4.0$ | 0.570±0.067 | 0.930±0.056 | 1.413±0.060 | 1.147 | 0.835±0.034 | 10.80±3.20 | 45.00±5.10 | 0.486 |
| | $4.0 < y < 4.5$ | 0.600±0.085 | 0.990±0.065 | 1.253±0.067 | 1.051 | 0.805±0.027 | 10.00±3.10 | 37.00±4.70 | 0.268 |
| Figure 2(a) | $2.0 < y < 2.5$ | 0.870±0.064 | 1.200±0.058 | 1.950±0.117 | 1.196 | 0.823±0.043 | 13.00±3.57 | 57.80±6.40 | 1.478 |
| | $2.5 < y < 3.0$ | 0.837±0.053 | 1.137±0.053 | 1.852±0.130 | 1.114 | 0.830±0.057 | 12.50±3.05 | 56.00±6.10 | 1.880 |
| | $3.0 < y < 3.5$ | 0.883±0.057 | 1.145±0.055 | 1.868±0.132 | 0.982 | 0.858±0.068 | 12.80±3.12 | 54.80±6.10 | 1.930 |
| | $3.5 < y < 4.0$ | 0.825±0.045 | 1.068±0.062 | 1.693±0.127 | 1.118 | 0.873±0.057 | 12.80±3.07 | 55.30±6.20 | 2.041 |
| | $4.0 < y < 4.5$ | 0.838±0.065 | 1.066±0.057 | 1.478±0.142 | 1.192 | 0.853±0.047 | 11.80±2.98 | 43.00±5.90 | 3.430 |
| | $2.0 < y < 2.5$ | 0.788±0.063 | 1.372±0.087 | 2.420±0.190 | 1.545 | 0.757±0.047 | 15.60±3.87 | 75.50±8.20 | 0.763 |
| | $2.5 < y < 3.0$ | 0.644±0.056 | 1.218±0.092 | 2.080±0.217 | 1.787 | 0.725±0.045 | 14.70±3.73 | 70.00±7.50 | 1.226 |
| Figure 2(b) | $3.0 < y < 3.5$ | 0.795±0.078 | 1.300±0.085 | 2.183±0.204 | 1.279 | 0.782±0.053 | 15.00±3.75 | 70.60±7.70 | 0.831 |
| | $3.5 < y < 4.0$ | 0.827±0.053 | 1.295±0.075 | 2.120±0.185 | 1.363 | 0.803±0.049 | 14.60±3.65 | 66.00±7.30 | 1.065 |
| | $4.0 < y < 4.5$ | 0.810±0.057 | 1.220±0.083 | 1.853±0.177 | 1.733 | 0.822±0.052 | 13.80±3.58 | 58.30±7.10 | 1.342 |
| Figure 3(a) | $2.0 < y < 2.5$ | 0.802±0.188 | 2.560±0.087 | 3.030±0.190 | 0.945 | 0.510±0.047 | 30.00±4.80 | 132.00±12.00 | 0.230 |
| | $2.5 < y < 3.0$ | 0.830±0.155 | 2.490±0.073 | 2.740±0.178 | 1.523 | 0.524±0.053 | 32.00±4.80 | 120.00±11.70 | 0.268 |
| | $3.0 < y < 3.5$ | 0.705±0.165 | 2.360±0.077 | 2.687±0.173 | 1.653 | 0.553±0.057 | 33.00±5.20 | 115.00±11.30 | 0.303 |
| | $3.5 < y < 4.0$ | 0.635±0.157 | 2.276±0.084 | 2.652±0.168 | 1.099 | 0.527±0.050 | 28.00±4.00 | 107.00±10.80 | 0.307 |
| | $4.0 < y < 4.5$ | 0.771±0.169 | 1.987±0.078 | 2.292±0.153 | 8.640 | 0.550±0.064 | 26.00±3.80 | 87.00±10.10 | 4.617 |
| Figure 3(b) | $2.0 < y < 2.5$ | 0.513±0.137 | 2.836±0.134 | 2.900±0.157 | 2.844 | 0.530±0.057 | 42.00±7.60 | 140.00±22.00 | 1.192 |
| | $2.5 < y < 3.0$ | 0.614±0.156 | 2.463±0.129 | 3.537±0.123 | 1.791 | 0.517±0.052 | 36.00±7.30 | 145.00±25.00 | 0.732 |
| | $3.0 < y < 3.5$ | 0.570±0.135 | 2.675±0.135 | 3.238±0.137 | 1.715 | 0.503±0.049 | 33.00±7.30 | 150.00±25.00 | 0.726 |
| | $3.5 < y < 4.0$ | 0.587±0.143 | 2.587±0.138 | 2.620±0.145 | 3.942 | 0.552±0.062 | 37.00±7.5 | 117.00±20.00 | 2.540 |
| | $4.0 < y < 4.5$ | 0.535±0.195 | 2.782±0.128 | 3.583±0.237 | 1.970 | 0.510±0.054 | 48.00±7.80 | 175.00±29.00 | 2.163 |
| Figure 3(c) | $2.0 < y < 2.5$ | 0.555±0.148 | 3.342±0.145 | 3.583±0.207 | 4.141 | 0.447±0.078 | 43.00±9.00 | 185.00±28.00 | 3.373 |
| | $2.5 < y < 3.0$ | 0.783±0.187 | 3.107±0.137 | 3.506±0.198 | 2.554 | 0.493±0.059 | 45.00±7.00 | 180.00±21.00 | 1.594 |
| | $3.0 < y < 3.5$ | 0.603±0.152 | 2.886±0.132 | 3.583±0.212 | 3.949 | 0.402±0.062 | 34.00±8.00 | 140.00±19.00 | 2.585 |
| | $3.5 < y < 4.0$ | 0.647±0.160 | 2.285±0.107 | 3.350±0.205 | 13.242 | 0.516±0.057 | 30.00±7.50 | 150.00±20.00 | 8.291 |
| | $4.0 < y < 4.5$ | 0.533±0.133 | 3.012±0.153 | 3.627±0.223 | 11.216 | 0.423±0.083 | 40.00±10.70 | 170.00±25.00 | 11.114 |
| Figure 4(a) | $1.5 < y < 2.0$ | 0.797±0.133 | 1.285±0.105 | 1.768±0.242 | 3.624 | 0.687±0.076 | 12.60±3.25 | 52.00±5.70 | 0.472 |
| | $2.0 < y < 2.5$ | 0.802±0.148 | 1.356±0.094 | 1.752±0.216 | 2.201 | 0.692±0.052 | 13.10±3.30 | 53.10±5.80 | 0.614 |
| | $2.5 < y < 3.0$ | 0.873±0.097 | 1.322±0.108 | 1.687±0.198 | 2.842 | 0.708±0.067 | 12.40±3.23 | 52.10±5.80 | 0.243 |
| | $3.0 < y < 3.5$ | 0.835±0.085 | 1.280±0.107 | 1.352±0.255 | 3.825 | 0.737±0.054 | 12.00±3.05 | 47.00±5.40 | 0.989 |
| | $3.5 < y < 4.0$ | 0.870±0.090 | 1.195±0.095 | 1.342±0.188 | 3.814 | 0.726±0.054 | 10.80±3.10 | 43.00±5.20 | 1.383 |
| Figure 4(b) | $1.5 < y < 2.0$ | 0.812±0.158 | 1.656±0.084 | 1.830±0.190 | 0.918 | 0.518±0.065 | 12.60±3.70 | 57.50±5.50 | 1.921 |
| | $2.0 < y < 2.5$ | 0.684±0.136 | 1.363±0.082 | 2.497±0.203 | 1.309 | 0.543±0.060 | 12.60±3.57 | 57.60±5.80 | 1.807 |
| | $2.5 < y < 3.0$ | 0.850±0.107 | 1.405±0.078 | 2.293±0.238 | 4.805 | 0.634±0.067 | 12.30±3.42 | 56.40±5.30 | 1.401 |
| | $3.0 < y < 3.5$ | 0.847±0.123 | 1.285±0.080 | 2.050±0.255 | 9.194 | 0.620±0.070 | 12.20±3.37 | 50.60±5.00 | 2.658 |
| | $3.5 < y < 4.0$ | 0.835±0.130 | 1.282±0.065 | 2.261±0.273 | 9.018 | 0.657±0.065 | 11.70±3.15 | 50.30±5.10 | 4.329 |
| Figure 5 | $C = 0-20\%$ | 0.702±0.158 | 0.937±0.060 | 0.960±0.195 | 0.778 | 0.567±0.063 | 5.00±0.170 | 17.00±2.30 | 0.012 |
| | $C = 20-40\%$ | 0.982±0.113 | 1.000±0.102 | 1.200±0.227 | 1.038 | 0.512±0.055 | 5.70±0.183 | 18.60±3.10 | 1.428 |
| | $C = 40-60\%$ | 0.823±0.083 | 1.020±0.087 | 1.200±0.253 | 1.163 | 0.518±0.047 | 5.80±0.180 | 19.40±3.42 | 1.274 |

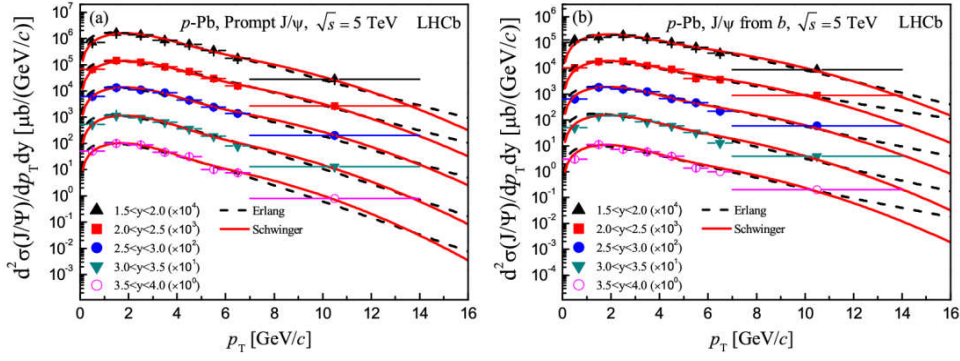

Figure 4: The same as Figure 2 but showing the results in different rapidity ranges in $p$-Pb collisions at $\sqrt{s_{NN}} = 5$ TeV. The data are taken from ref. [20].

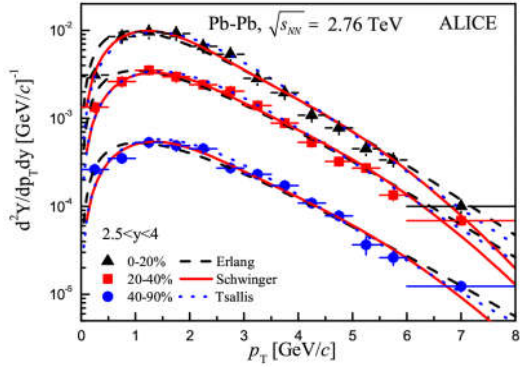

Figure 5: Transverse momentum spectra of $J/\psi$ produced in three centrality classes in Pb-Pb collisions at $\sqrt{s_{NN}} = 2.76$ TeV. The symbols represent the experimental data of the ALICE Collaboration [21]. The dashed, solid, and dotted curves are our results calculated by using the two-component Erlang distribution, the two-component Schwinger mechanism, and the Tsallis statistics, respectively.

The transverse momentum spectra, $d^2Y/dp_T dy$, of $J/\psi$ produced in three centrality classes and $2.5 < y < 4$ in Pb-Pb collisions at $\sqrt{s_{NN}} = 2.76$ TeV are given in Figure 5, where $Y$ denotes the yield. The symbols represent the experimental data of the ALICE collaboration [21]. The dashed and solid curves are our results calculated by using the two-component Erlang distribution and the two-component Schwinger mechanism respectively. The values of free parameters and $\chi^2/dof$ are listed in Table 1. Similar conclusion obtained from Figure 1 can be obtained from Figure 5.

To give a comparison of fit quality with some other approach, as an example, we show the result of the Tsallis statistics [7-11] in Figure 5 by the dotted curves. A simplified form of the Tsallis transverse momentum distribution

$$f(p_T) = C_T p_T \sqrt{p_T^2 + m_0^2} \left[1 + \frac{q-1}{T}\left(\sqrt{p_T^2 + m_0^2} - \mu\right)\right]^{-q/(q-1)} \quad (7)$$

is used, where $T$ is the temperature parameter, $q$ is the entropy index, $\mu$ is the chemical potential which is regarded as 0 at the LHC, and $C_T$ is the normalization constant. In Eq. (6), the effect of longitudinal motion is subtracted by using $y = 0$ to obtain the temperature parameter as accurately as possible. For the centrality from

0-20% to 40-90%, the temperature parameter is taken to be 0.285±0.008, 0.273±0.006, and 0.264±0.009 GeV, the entropy index is taken to be 1.067±0.003, 1.080±0.004, and 1.085±0.007, with $\chi^2/dof$ to be 0.969, 1.290, and 1.781, respectively. One can see the compatibility of the three approaches, which shows the multiformity of fit functions.

In the above fit to the experimental data of LHCb and ALICE Collaborations, the uncorrelated and correlated uncertainties in experimental data are together included in the calculation of $\chi^2/dof$ by using the quadratic sums. No matter for the part of uncorrelated or correlated uncertainty, even for the part of correlated, especially multiplicative, common for all bins uncertainties, we just use the experimental values directly.

To see clearly the relationships between parameters and rapidity, parameters and centrality, as we as parameters and others, we plot the values listed in Table 1 in Figures 6-11, which correspond to the relationships related to parameters $k_{E1}$, $\langle p_t \rangle_1$, $\langle p_t \rangle_2$, $k_{S1}$, $\kappa_1$, and $\kappa_2$, respectively. In these figures, the symbols and lines are parameter values and fitting lines respectively. The intercepts, slopes, and $\chi^2/dof$ related to the lines in Figures 6-11 are listed in Table 2. In the error range, $k_{E1}$ and $k_{S1}$ do not show obvious change with rapidity, centrality, energy, and size. $\langle p_t \rangle_1$, $\langle p_t \rangle_2$, $\kappa_1$, and $\kappa_2$ do not show obvious change with rapidity and centrality, or they decrease slightly with increases of rapidity and centrality.

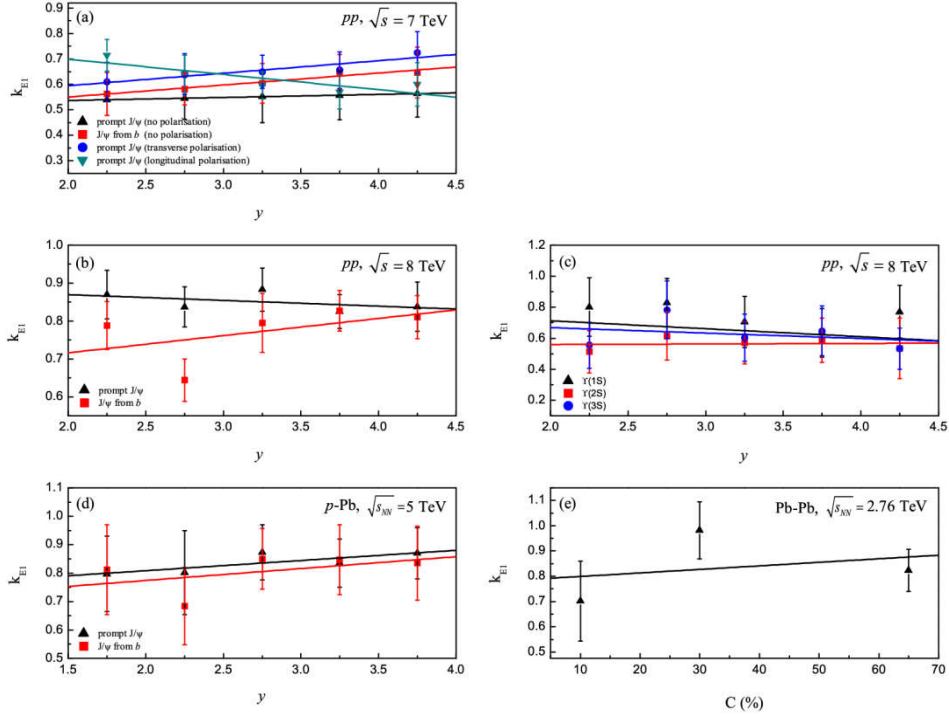

Figure 6: Dependences of contribution ratio $k_{E1}$ for (a) $J/\psi$ on $y$ in $pp$ collision at $\sqrt{s}=7$ TeV, (b) $J/\psi$ on $y$ in $pp$ collision at $\sqrt{s}=8$ TeV, (c) $\Upsilon$ on $y$ in $pp$ collision at $\sqrt{s}=8$ TeV, (d) $J/\psi$ on $y$ in $p$-Pb collisions at $\sqrt{s}=5$ TeV, and (e) $J/\psi$ on $C$ in Pb-Pb collisions at $\sqrt{s_{NN}}=2.76$ TeV. The symbols are quoted in Table 1 and the lines are our fitting results.

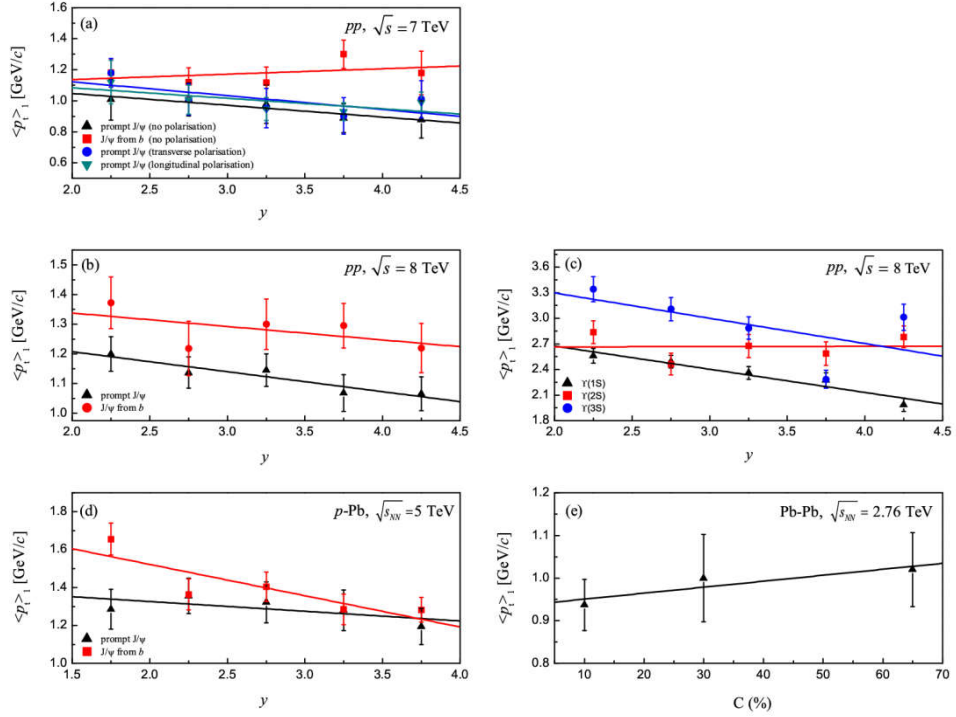

Figure 7: The same as Figure 6 but showing the results for $\langle p_t \rangle_1$.

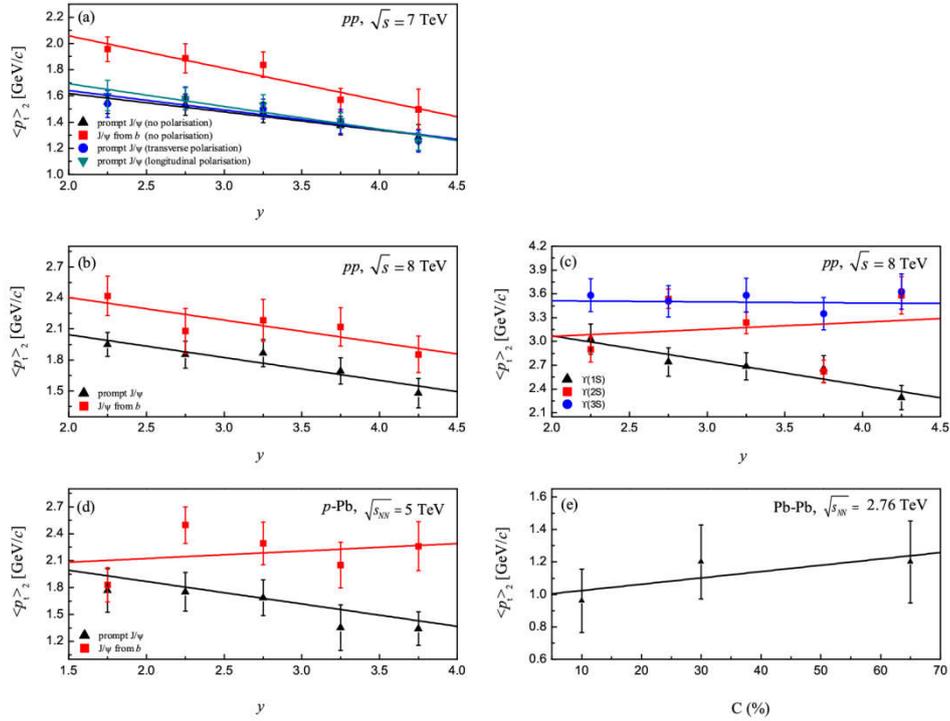

Figure 8: The same as Figure 6 but showing the results for $\langle p_t \rangle_2$.

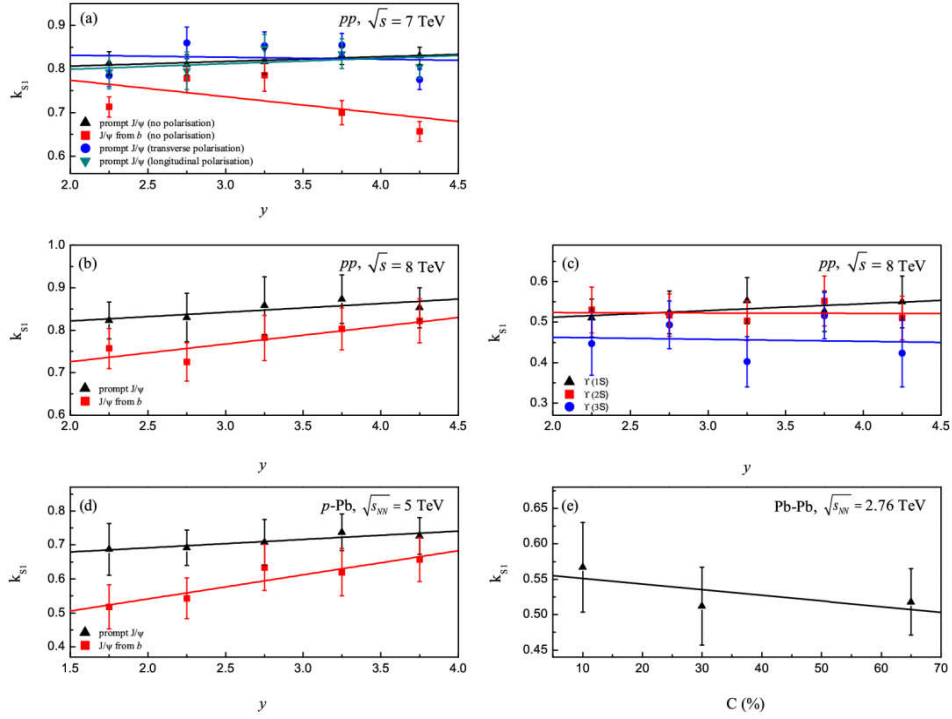

Figure 9: The same as Figure 6 but showing the results for $k_{S1}$.

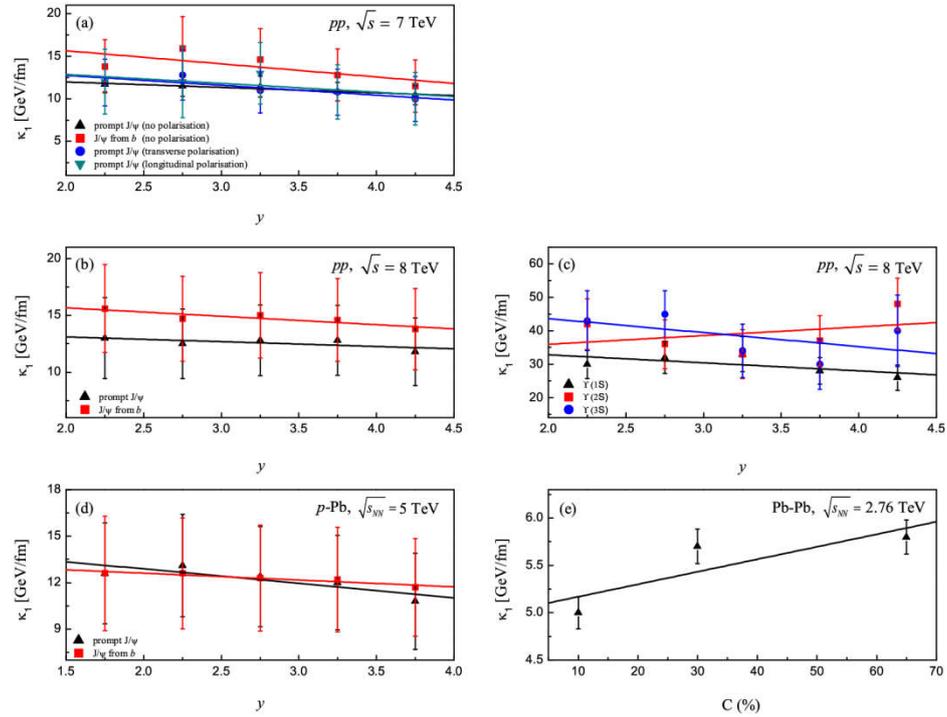

Figure 10: The same as Figure 6 but showing the results for $\kappa_1$.

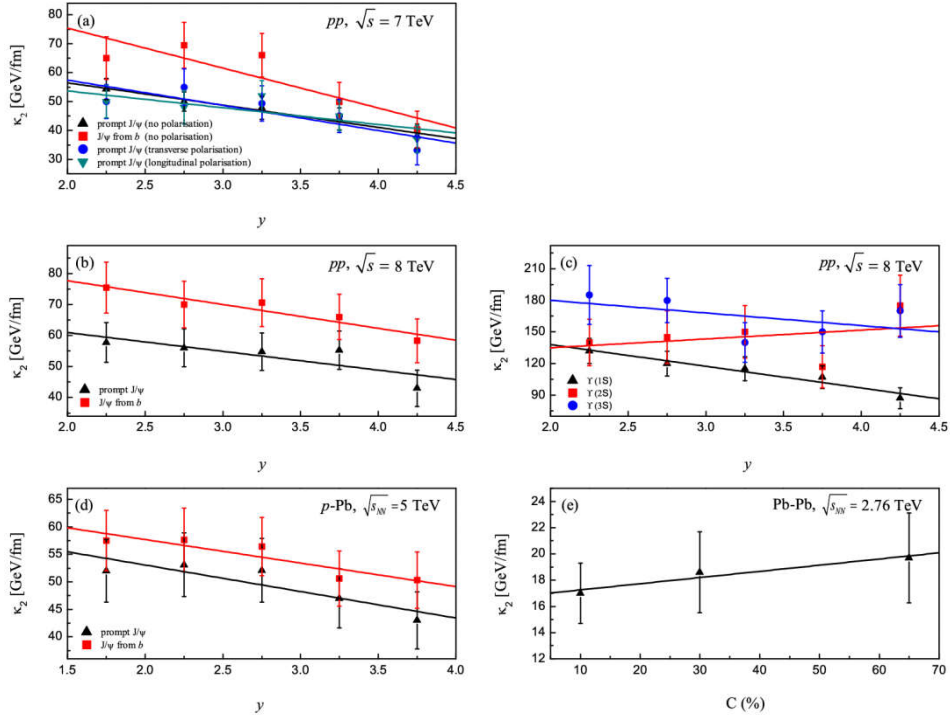

Figure 11: The same as Figure 6 but showing the results for $\kappa_2$.

Table 2: Values of intercepts, slopes, and $\chi^2/dof$ corresponding to the lines in Figures 6-11.

| Figure | Type | Intercept | Slope | $\chi^2/dof$ |
|---|---|---|---|---|
| Figure 6(a) | Prompt $J/\psi$ (no polarisation) | 0.513±0.001 | 0.012±0.001 | 0.001 |
| | $J/\psi$ from $b$ (no polarisation) | 0.456±0.021 | 0.047±0.006 | 0.003 |
| | Prompt $J/\psi$ (transverse polarisation) | 0.496±0.035 | 0.049±0.011 | 0.007 |
| | Prompt $J/\psi$ (longitudinal polarisation) | 0.819±0.065 | -0.060±0.020 | 0.030 |
| Figure 6(b) | Prompt $J/\psi$ | 0.900±0.045 | -0.015±0.014 | 0.013 |
| | $J/\psi$ from $b$ | 0.625±0.135 | 0.045±0.041 | 0.139 |
| Figure 6(c) | $\Upsilon(1S)$ | 0.916±0.142 | -0.051±0.043 | 0.053 |
| | $\Upsilon(2S)$ | 0.553±0.085 | 0.003±0.025 | 0.025 |
| | $\Upsilon(3S)$ | 0.741±0.200 | -0.036±0.060 | 0.111 |
| Figure 6(d) | Prompt $J/\psi$ | 0.737±0.040 | 0.036±0.014 | 0.008 |
| | $J/\psi$ from $b$ | 0.691±0.110 | 0.042±0.039 | 0.051 |
| Figure 6(e) | Inclusive | 0.785±0.143 | 0.001±0.003 | 0.344 |
| Figure 7(a) | Prompt $J/\psi$ (no polarisation) | 1.200±0.049 | -0.076±0.015 | 0.007 |
| | $J/\psi$ from $b$ (no polarisation) | 1.067±0.144 | 0.035±0.043 | 0.054 |
| | Prompt $J/\psi$ (transverse polarisation) | 1.301±0.163 | -0.089±0.049 | 0.070 |
| | Prompt $J/\psi$ (longitudinal polarisation) | 1.219±0.110 | -0.068±0.033 | 0.047 |
| Figure 7(b) | Prompt $J/\psi$ | 1.342±0.041 | -0.068±0.012 | 0.008 |
| | $J/\psi$ from $b$ | 1.429±0.112 | -0.045±0.034 | 0.034 |
| Figure 7(c) | $\Upsilon(1S)$ | 3.219±0.128 | -0.272±0.039 | 0.028 |
| | $\Upsilon(2S)$ | 2.658±0.315 | 0.003±0.095 | 0.087 |
| | $\Upsilon(3S)$ | 3.890±0.670 | -0.296±0.202 | 0.449 |
| Figure 7(d) | Prompt $J/\psi$ | 1.428±0.080 | -0.051±0.028 | 0.021 |
| | $J/\psi$ from $b$ | 1.853±0.144 | -0.165±0.051 | 0.075 |

| Figure | | | | |
|---|---|---|---|---|
| Figure 7(e) | Inclusive | 0.937±0.020 | 0.001±0.001 | 0.008 |
| Figure 8(a) | Prompt $J/\psi$ (no polarisation) | 1.898±0.054 | -0.140±0.016 | 0.007 |
| | $J/\psi$ from $b$ (no polarisation) | 2.552±0.119 | -0.247±0.036 | 0.025 |
| | Prompt $J/\psi$ (transverse polarisation) | 1.938±0.116 | -0.148±0.035 | 0.031 |
| | Prompt $J/\psi$ (longitudinal polarisation) | 2.039±0.096 | -0.173±0.029 | 0.025 |
| Figure 8(b) | Prompt $J/\psi$ | 2.485±0.142 | -0.221±0.043 | 0.026 |
| | $J/\psi$ from $b$ | 2.842±0.227 | -0.219±0.068 | 0.036 |
| Figure 8(c) | $\Upsilon(1S)$ | 3.697±0.191 | -0.313±0.057 | 0.025 |
| | $\Upsilon(2S)$ | 2.884±0.858 | 0.090±0.258 | 0.511 |
| | $\Upsilon(3S)$ | 3.574±0.229 | -0.014±0.069 | 0.022 |
| Figure 8(d) | Prompt $J/\psi$ | 2.269±0.151 | -0.250±0.053 | 0.027 |
| | $J/\psi$ from $b$ | 1.958±0.442 | 0.083±0.156 | 0.180 |
| Figure 8(e) | Inclusive | 0.985±0.092 | 0.004±0.002 | 0.062 |
| Figure 9(a) | Prompt $J/\psi$ (no polarisation) | 0.785±0.009 | 0.011±0.003 | 0.001 |
| | $J/\psi$ from $b$ (no polarisation) | 0.850±0.096 | -0.038±0.029 | 0.132 |
| | Prompt $J/\psi$ (transverse polarisation) | 0.841±0.087 | -0.005±0.026 | 0.108 |
| | Prompt $J/\psi$ (longitudinal polarisation) | 0.774±0.047 | 0.013±0.014 | 0.025 |
| Figure 9(b) | Prompt $J/\psi$ | 0.780±0.027 | 0.021±0.008 | 0.005 |
| | $J/\psi$ from $b$ | 0.643±0.041 | 0.042±0.012 | 0.015 |
| Figure 9(c) | $\Upsilon(1S)$ | 0.479±0.027 | 0.017±0.008 | 0.007 |
| | $\Upsilon(2S)$ | 0.526±0.041 | -0.001±0.012 | 0.016 |
| | $\Upsilon(3S)$ | 0.472±0.100 | -0.005±0.030 | 0.106 |
| Figure 9(d) | Prompt $J/\psi$ | 0.642±0.016 | 0.025±0.006 | 0.003 |
| | $J/\psi$ from $b$ | 0.399±0.040 | 0.071±0.014 | 0.017 |
| Figure 9(e) | Inclusive | 0.559±0.023 | -0.001±0.001 | 0.031 |
| Figure 10(a) | Prompt $J/\psi$ (no polarisation) | 13.254±0.305 | -0.638±0.092 | 0.002 |
| | $J/\psi$ from $b$ (no polarisation) | 18.725±2.440 | -1.540±0.733 | 0.037 |
| | Prompt $J/\psi$ (transverse polarisation) | 15.070±1.188 | -1.160±0.357 | 0.012 |
| | Prompt $J/\psi$ (longitudinal polarisation) | 14.940±1.750 | -1.040±0.526 | 0.021 |
| Figure 10(b) | Prompt $J/\psi$ | 13.945±0.703 | -0.420±0.211 | 0.004 |
| | $J/\psi$ from $b$ | 17.145±0.616 | -0.740±0.185 | 0.002 |
| Figure 10(c) | $\Upsilon(1S)$ | 37.600±4.512 | -2.400±1.357 | 0.042 |
| | $\Upsilon(2S)$ | 30.750±11.613 | 2.600±3.491 | 0.137 |
| | $\Upsilon(3S)$ | 52.050±11.186 | -4.200±3.363 | 0.130 |
| Figure 10(d) | Prompt $J/\psi$ | 14.765±0.803 | -0.840±0.283 | 0.007 |
| | $J/\psi$ from $b$ | 13.490±0.227 | -0.440±0.080 | 0.001 |
| Figure 10(e) | Inclusive | 5.037±0.247 | 0.013±0.006 | 0.111 |
| Figure 11(a) | Prompt $J/\psi$ (no polarisation) | 71.714±2.053 | -7.720±0.617 | 0.007 |
| | $J/\psi$ from $b$ (no polarisation) | 103.035±12.805 | -13.820±3.850 | 0.110 |
| | Prompt $J/\psi$ (transverse polarisation) | 74.840±9.489 | -8.720±2.853 | 0.107 |
| | Prompt $J/\psi$ (longitudinal polarisation) | 65.250±7.664 | -5.800±2.304 | 0.075 |
| Figure 11(b) | Prompt $J/\psi$ | 73.075±7.290 | -6.060±2.192 | 0.053 |
| | $J/\psi$ from $b$ | 93.040±4.414 | -7.680±1.327 | 0.012 |
| Figure 11(c) | $\Upsilon(1S)$ | 179.150±8.275 | -20.600±2.488 | 0.019 |
| | $\Upsilon(2S)$ | 118.100±41.500 | 8.400±12.477 | 0.185 |
| | $\Upsilon(3S)$ | 204.000±35.512 | -12.000±10.677 | 0.121 |
| Figure 11(d) | Prompt $J/\psi$ | 62.695±3.633 | -4.820±1.280 | 0.019 |
| | $J/\psi$ from $b$ | 66.250±2.734 | -4.280±0.963 | 0.011 |
| Figure 11(e) | Inclusive | 16.785±0.374 | 0.047±0.009 | 0.005 |

## 4. Discussions

To discuss on the Schwinger mechanism, if the charmonium ($c\bar{c}$) or bottomonium ($b\bar{b}$) can be produced in the collision, the minimum distance between the two partons, for the $i$th component, is [17]

$$L_{\min,i} = 2m_T/\kappa_i = 2\sqrt{p_T^2 + m_0^2}/\kappa_i, \tag{8}$$

where $m_0$ is the mass of produced charmed or bottom quark, but not that of the participant parton. The mean minimum distance

$$\langle L_{\min,i} \rangle = 2\langle m_T \rangle/\kappa_i = 2\langle \sqrt{p_T^2 + m_0^2} \rangle/\kappa_i. \tag{9}$$

The minimal minimum distance

$$(L_{\min,i})_{\min} = 2m_0/\kappa_i. \tag{10}$$

We see that the string tension is an important parameter which is related to the minimum distance, the mean minimum distance, and the minimal minimum distance. Correspondingly, the distribution of the minimum distance can be obtained by $(1/N)(dN/dL_{\min,i})$, where $N$ denotes the number of charmoniums.

Further, if the produced charmed or bottom quark stays at haphazard at the middle between the two participant partons, the maximum potential energy of the charmed quark staying in the colour field of the two partons is

$$V_{\max,i} = -\kappa_i L_{\min,i}/2 = -m_T = -\sqrt{p_T^2 + m_0^2}. \tag{11}$$

The mean maximum potential energy

$$\langle V_{\max,i} \rangle = -\langle \kappa_i L_{\min,i}/2 \rangle = -\langle m_T \rangle = -\langle \sqrt{p_T^2 + m_0^2} \rangle. \tag{12}$$

The maximal maximum potential energy

$$(V_{\max,i})_{\max} = -(\kappa_i L_{\min,i}/2)_{\max} = -m_0. \tag{13}$$

Correspondingly, the distribution of the maximum potential energy can be obtained by $(1/N)(dN/dV_{\max,i})$.

From Table 1 we see that the values of $\langle p_t \rangle_1$, $\langle p_t \rangle_2$, $\kappa_1$, and $\kappa_2$ in the collisions at the LHC are large, which render that the interactions between partons are violent and the minimal minimum distance between the two interacting partons is small. According to $\kappa_2$, the minimal minimum distance is ~0.03-0.06 fm which is a few percent of nucleon size. We believe that, in the collisions at the LHC, nucleons penetrate through each other totally. Partons in projectile nucleon have large probability to close exceedingly to partons in target nucleons. At the same time, both the contribution ratios of the two second components in the two calculation methods are large and cannot be neglected.

## 5. Conclusions

From the above discussions, we obtain the following conclusions:

(a) The transverse momentum spectra of $J/\psi$ and $\Upsilon$ mesons produced in high energy collisions are described by using both the two-component Erlang distribution and the two-component Schwinger mechanism. The results obtained by the two methods are compared and found to be in agreement with the experimental data of *pp*, *p*-Pb, and Pb-Pb collisions at the LHC.

(b) The related parameters such as the mean transverse momentum $\langle p_t \rangle_1$ ($\langle p_t \rangle_2$) contributed by each parton in the first (second) component in the two-component Erlang distribution, and the string tension $\kappa_1$ ($\kappa_2$) between two partons in the first (second) component in the two-component Schwinger mechanism are extracted. At the same time, the contribution ratios $k_{E1}$ and $k_{S1}$ of the two first components in the two methods are obtained.

(c) In the error range, $k_{E1}$ and $k_{S1}$ do not show obvious change with rapidity, centrality, energy, and size. $\langle p_t \rangle_1$, $\langle p_t \rangle_2$, $\kappa_1$, and $\kappa_2$ do not show obvious change with rapidity and centrality, or they decrease slightly with increases of rapidity and centrality. Both the contribution ratios of the two second components cannot be neglected. The minimal minimum distance between interacting partons is ~0.03-0.06 fm which is a few percent of nucleon size.

**Conflict of Interests**

The authors declare that there is no conflict of interests regarding the publication of this paper.

**Acknowledgments**

This work was supported by the National Natural Science Foundation of China under Grant No. 11575103.

**References**

[1] N. Itoh, "Hydrostatic equilibrium of hypothetical quark stars," *Progress in Theoretical Physics*, vol. 44, no. 1, pp. 291–292, 1970.

[2] J. R. Nix, "Theory of high-energy heavy-ion collisions," *Progress in Particle and Nuclear Physics*, vol. 2, pp. 237–284, 1979.

[3] M. Stephanov, K. Rajagopal, and E. Shuryak, "Signatures of the tricritical point in QCD," *Physical Review Letters*, vol. 81, no. 22, pp. 4816–4819, 1998.

[4] S. Abreu, S. V. Akkelin, J. Alam et al., "Heavy ion collisions at the LHC—last call for predictions," *Journal of Physics G*, vol. 35, no. 5, Article ID 054001, 185 pages, 2008.

[5] U. Heinz, "Concepts of heavy-ion physics," in *Proceedings of the 2nd CERN—Latin American School of High Energy Physics*, Lecture Notes for Lectures, San Miguel Regla, Mexico, June 2003.

[6] P. Z. Ning, L. Li, and D. F. Min, *Foundation of Nuclear Physics: Nucleons and Nuclei*, Higher Education Press, Beijing, China, 2003.

[7] C. Tsallis, "Possible generalization of Boltzmann-Gibbs statistics," *Journal of Statistical Physics*, vol. 52, no. 1-2, pp. 479–487, 1988.

[8] C. Tsallis, "Nonadditive entropy and nonextensive statistical mechanics—an overview after 20 years," *Brazilian Journal of Physics*, vol. 39, no. 2, pp. 337–356, 2009.

[9] T. S. Biró, G. Purcsel, K. Ürmössy, "Non-extensive approach to quark matter," *European Physical Journal A*, vol. 40, no. 3, pp. 325–340, 2009.

[10] G. Wilk and Z. Włodarczyk, "Multiplicity fluctuations due to the temperature

fluctuations in high-energy nuclear collisions," *Physical Review C*, vol. 79, no. 5, Article ID 054903, 10 pages, 2009.

[11] J. Cleymans and D. Worku, "Relativistic thermodynamics: transverse momentum distributions in high-energy physics," *European Physical Journal A*, vol. 48, Article ID 160, 8 pages, 2012.

[12] H.-R. Wei, Y.-H. Chen, L.-N. Gao, and F.-H. Liu, "Comparing multicomponent Erlang distribution and Levy distribution of particle transverse momentum," *Advances in High Energy Physics*, vol. 2014, Article ID 782631, 16 pages, 2014.

[13] J. Schwinger, "On gauge invariance and vacuum polarization," *Physical Review*, vol. 82, no. 5, pp. 664–679, 1951.

[14] R.-C. Wang and C.-Y. Wong, "Finite-size effect in the Schwinger particle-production mechanism," *Physical Review D*, vol. 38, no. 1, pp. 348–359, 1988.

[15] P. Braun-Munzinger, K. Redlich, and J. Stachel, "Particle production in heavy ion collisions," in *Quark-Gluon Plasma 3*, eds. R. C. Hwa and X.-N. Wang, World Scientific, Singapore, 2004

[16] L. G. Gutay, A. S. Hirsch, C. Pajares, R. P. Scharenberg, and B. K. Srivastava, "De-confinement in small systems: clustering of color source in high multiplicity $\bar{p}p$ collisions at $\sqrt{s}=1.8$ TeV," http://arxiv.org/abs/1504.08270.

[17] C.-Y. Wong, *Introduction to High Energy Heavy Ion Collisions*, World Scientific, Singapore, 1994.

[18] LHCb Collaboration, R. Aaij, B. Adeva, M. Adinolfi et al., "Measurement of $J/\psi$ production in *pp* collisions at $\sqrt{s}=7$ TeV," *European Physical Journal C*, vol. 71, no. 5, Article ID 1645, 17 pages, 2011.

[19] LHCb Collaboration, R. Aaij, C. A. Abellan, B. Adeva et al., "Production of $J/\psi$ and $\Upsilon$ mesons in *pp* collisions at $\sqrt{s}=8$ TeV," *Journal of High Energy Physics*, vol. 1306, Article ID 064, 30 pages, 2013.

[20] LHCb Collaboration, R. Aaij, B. Adeva, M. Adinolfi et al., "Study of $J/\psi$ production and cold nuclear matter effects in *p*-Pb collisions at $\sqrt{s_{NN}}=5$ TeV," *Journal of High Energy Physics*, vol. 1402, Article ID 072, 22 pages, 2014.

[21] ALICE Collaboration, J. Adam, D. Adamová, M. M. Aggarwal et al., "Differential studies of inclusive $J/\psi$ and $\psi(2S)$ production at forward rapidity in Pb-Pb collisions at $\sqrt{s_{NN}}=2.76$ TeV," http://arxiv.org/abs/1506.08804, *Journal of High Energy Physics*, submitted.